\documentclass{elsart}

\usepackage{graphics}
\usepackage{color} 
\usepackage[normalem]{ulem}
\usepackage{colortbl}

\newcommand{\be}{\begin{equation}}
\newcommand{\ee}{\end{equation}}
\newcommand{\bea}{\begin{eqnarray}}
\newcommand{\eea}{\end{eqnarray}}

\begin{document}
\begin{frontmatter}

\title{New aspects of the continuous phase transition in the
scalar noise model (SNM) of collective motion}

\author{M\'at\'e Nagy$^1$, Istv\'an Daruka$^2$ and Tam\'as Vicsek$^{1, 2}$}

\address{$^1$ Biological Physics Research Group of HAS,
 P\'azm\'any P.\ stny.\ 1A, H-1117 Budapest, Hungary}

\address{$^2$ Department of Biological Physics, E\"otv\"os University,
 P\'azm\'any P.\ stny.\ 1A, H-1117 Budapest, Hungary.}


\begin{abstract}
In this paper we present our detailed investigations on the
nature of the phase transition in the scalar noise model (SNM)
of collective motion. Our results confirm the
original findings of Ref. \cite{vicsek} that the
disorder-order transition in the SNM is a continuous, second order 
phase transition for small particle velocities ($v\leq 0.1$). 
However, for large velocities ($v\geq 0.3$) we find a strong anisotropy
in the particle diffusion in contrast with the isotropic diffusion
for small velocities. The interplay between the anisotropic diffusion and
the periodic boundary conditions leads to an artificial symmetry breaking
of the solutions (directionally quantized density waves) and a consequent 
first order transition like behavior. 
Thus, it is not possible to draw any conclusion about the 
physical behavior in the large particle velocity regime of the SNM.

\end{abstract}

\end{frontmatter}

\section{Introduction} 

In the recent years there has been a high interest in the study and modeling 
of collective behavior of living systems. Among the diverse and startling
features of the collective behavior (e.g., synchronization \cite{sync}),
collective motion observed in bird flocks \cite{birds},
fish schools \cite{fish}, insect swarms \cite{insect}, and bacteria aggregates 
\cite{bacteria} is one
of the most profound manifestations. The powerful tools 
(scale invariance and renormalization) of statistical physics enable us to
effectively model and investigate the nature of motion in nonequilibrium 
many-particle systems. 
In particular, due to their strong analogy with living systems, self-propelled
particle models \cite{vicsek,czirok,csahok,tonertu1,tonertu2,tonertu3,gregoire} play a crucial role in understanding 
the key features of such biological systems. One such important aspect
of these models, observed also in bird flocks, is the onset of collective 
motion without a leader. Furthermore, self-propelled 
particle models exhibit a behavior analogous to the phase transitions in
equilibrium systems. It means that there is a kinetic phase transition from 
the disordered (high noise or temperature) state to an ordered (low noise or
temperature) state where all the particles move more or less in the same
direction \cite{vicsek}. In this paper we present our results on the
new aspects of the collective motion occurring in the scalar noise model
(SNM) of self-propelled particles.  

The model of Vicsek et al. \cite{vicsek}  was developed to study the
collective motion of self-propelled particles. The 
original model assumes a constant absolute particle velocity 
$v$ and includes a velocity direction averaging interaction
within a radius $R$.  Also, there is some random noise introduced
in the velocity update to mimic realistic situations (e.g. bird flight 
or bacteria motion). In particular, the model was implemented
on a square cell of linear size $L$ with periodic boundary
conditions. The density of a system with $N$ particles is defined
as $\rho = N/L^2$. 
The range of interaction was set to unity ($R$ = 1)
and the time step between two updates was chosen to be $\Delta t$ = 1. 
As initial condition, the particles were randomly distributed
in the cell, the particles had the same absolute velocity $v$
with randomly distributed directions $\theta$. 

The velocities \{{$\bf v_i$} \} of the particles were determined
simultaneously at each time step and the position of
the {\it i}th particle was updated according to 
\be
{\bf x_i}(t+\Delta t)= {\bf x_i}(t)+{\bf v_i}(t)\Delta t. 
\ee

The velocity {$\bf v_i$} of the {\it i}th particle was characterized
by its constant absolute value $v$ and its directional angle
$\theta$. The angle was updated as follows: 
\be 
\theta (t+\Delta t) = <\theta(t)>_R +\Delta\theta, 
\ee
where $<\theta(t)>_R$ denotes the average direction of the velocities
of particles (including particle {it i}) within the radius of interaction $R$. 
Furthermore, $\Delta\theta$ is a random number chosen with a uniform
probability from the interval $\eta[-\pi, \pi]$, where $\eta$ is the strength of the scalar noise .
This latter term, $\Delta\theta$ represents a scalar type of noise
in the system and therefore we refer to the above model as
the scalar noise model (SNM) of collective motion. 
This is important to point out here that this original model was
developed to describe the continuous motion of bacteria and/or birds.
To obtain a good approximation of this aim by the numerical implementation, 
we need to have many time steps performed before a significant change occurs 
in the neighborhood of the particles. This requirement can be 
expressed by the quantitative condition $v\Delta t \ll$ 1. This corresponds 
to the small velocity regime of the model and we expect proper
physical and biological features to be revealed in this limit.  
In order to characterize the collective behavior of the particles, 
the order parameter 
\be \varphi= \frac{1}{Nv}\vert \sum_{i=1}^N{\bf v_i}\vert
\ee
was introduced. Clearly, 
this parameter corresponds to the normalized average velocity
of the $N$ particles comprising the system. If the particles
move more or less randomly, the order parameter is approximately
zero, and if all the particles move in one direction, the order parameter
becomes unity.

\section{Kinetic Phase Transition} 

The physical behavior of the SNM in the small velocity regime 
was investigated from many aspects \cite{vicsek,czirok,tonertu3,aldana}. 
It has been
established that there is an ordering of particles as the noise
is decreased below a (particle density and velocity dependent) critical 
value. The original investigations of Ref. \cite{vicsek,czirok}
show that this order-disorder transition is a second order
phase transition. Furthermore, the order parameter $\varphi$ was found to satisfy 
the following scaling relations

\be 
\varphi \sim [\eta_c(\rho ) - \eta ]^\beta \hskip 0.7cm {\rm and} \hskip 0.7cm
\varphi \sim [\rho - \rho_c(\eta) ]^\delta, 
\ee
where $\beta$ = 0.45 $\pm$ 0.07 and $\delta$ = 0.35 $\pm$ 0.06 are critical exponents 
and $\eta_c(\rho )$ and $\rho_c(\eta)$
are the critical noise and density (for $L \to \infty$), respectively. 

Gr\'egoire and Chat\'e \cite{gregoire} argued that the second order nature of this
phase transition was due to some strong finite size effects and
they claimed that the phase transition was of first order in the SNM and the 
ordered phase is being a density wave.

We re-investigated their simulations using the same parameters (system size $L$ = 512, 
and particle velocity $v$ = 0.5) and boundary conditions (periodic) they implemented. 
In particular, we were interested in obtaining the probability distribution
function (PDF) of the order parameter $\varphi$ and also the pertaining Binder 
cumulant $G$. The Binder cumulant \cite{binder} defined as 
$G=1-<\varphi^4>/3<\varphi^2>^2$ measures
the fluctuations of the order parameter and is a good measure to distinguish
between first and second order phase transitions. In case of a first
order phase transition $G$ has a definite minimum, while for a second order
transition $G$ does not exhibit a characteristic minimum. 
Our new results, shown in Fig. 1,  were in complete disagreement with those of Fig. 2
of Ref. \cite{gregoire}. We found that the PDF was only one humped and
also, under these conditions we could not find well defined minimum in $G$. 
Our personal correspondence with Chat\'e and Gr\'egoire revealed the reason
of this discrepancy.  It turned out that Ref. \cite{gregoire} 
used a velocity updating rule that was different from that of the original paper 
of Vicsek et al. \cite{vicsek}. This difference was big enough to change the 
order of the phase transition derived from the numerical simulations.  
Furthermore, in Section 4 we demonstrate that that due to the presence of 
an inherent numerical artifact, it is not possible to give a physical
interpretation of the results in the large velocity regime ($v\geq 0.3$). 

\begin{figure}[t!]
\centerline{\resizebox{0.9\textwidth}{!}{\rotatebox{0}{\includegraphics{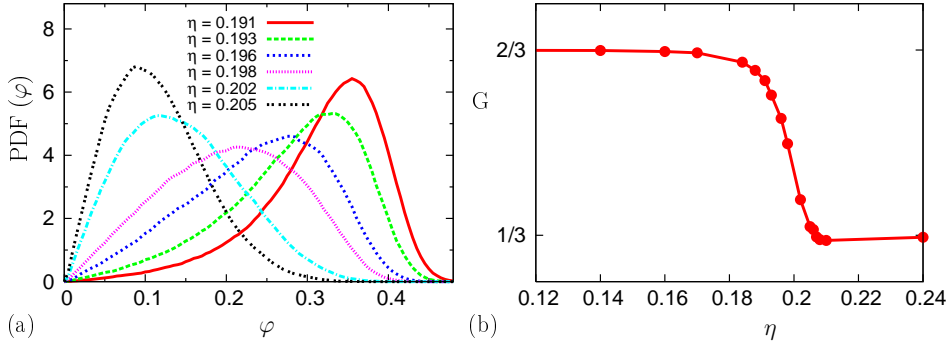}}}}
\caption{ (a) The probability distribution function (PDF) of the order parameter 
$\varphi$ (defined in the text) for noise
values $\eta$ around the critical point. The one humped curves indicate
a {\it second order phase transition}. (b) The Binder cumulant $G$ (defined
in the text) as a function of the noise. The smooth behavior of $G$ 
corroborates the second order nature of the disorder-order transition. 
The curves (a), (b) were
obtained for the same parameters as in Ref. \cite{gregoire}: 
linear system size $L$ = 512, particle density 
$\rho$ = 1/8, velocity $v$ = 0.5 and we used $\simeq300\tau$ MC simulation steps after observing 
the $\tau\simeq 10^5$ relaxation time. }
\label{fig:Pdf_v0.5}
\end{figure}

\section{The small velocity regime} 

The original model of Vicsek et al. was proposed to study the motion of bird flocks 
and/or bacterial colonies. The motion of the particles in such systems is   
quasi-continuous, i.e., usually the reaction time of the birds is significantly faster
than the characteristic time that is needed to travel through their interaction radius ($R$). 
This condition imposes the following constraint on the update time $\Delta t$ 
in the numerical simulations: $\Delta t \ll R/v$, where $v$ is the magnitude of
the particle velocity.  After fixing the interaction radius $R$ = 1 and the update time
$\Delta t$ = 1, the above condition becomes $v \ll$ 1. We refer to this velocity domain
as the {\it small velocity regime}. Our further investigations (discussed
below) showed that the small velocity regime actually holds for velocities $v\le$ 0.1.
There may be (relatively) rare situations when the large velocity regime 
($v\geq 0.3$) used by Gr\'egoire and Chat\'e \cite{gregoire} is a reasonable approximation 
of the flocking process (e.g., 'turbulent' escape motion of birds during the
attack of a predator), however, the physical justification of such situations 
is beyond the scope of the present
paper. In the large velocity regime Ref. \cite{gregoire} finds density waves in
the ordered state, objects that were not present in the simulations of Ref. \cite{vicsek}. 
We discuss the behavior of these planar waves occurring in the large velocity regime 
in the next section. 

Intrigued by the possibility of finding density waves,
we re-investigated the small velocity regime of \cite{vicsek} at larger system 
sizes and significantly longer simulation 
times. We carried out a series of runs for different velocities from $v$ = 0.01 to $v$ = 0.1.
Typical snapshots of the behavior are shown in Fig. 2. On can see isolated and 
uncorrelated, but coherently moving flocks in the system. The flocks have reached their steady
sizes. The nature of the disorder-order phase transition was characterized by the
probability distribution function (PDF) of the order parameter $\varphi$ (average particle
velocity). 

\begin{figure}[t!]
\centerline{\resizebox{0.4\textwidth}{!}{\rotatebox{0}{\includegraphics{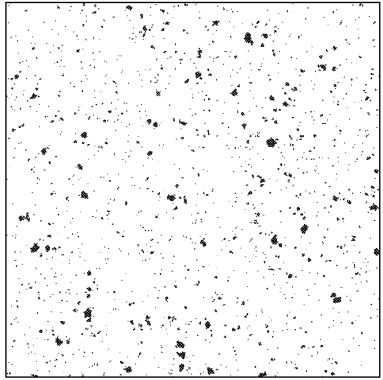}}}}
\caption{A typical snapshot of the system in the small velocity regime ($L$ = 512, $\rho$ = 1/8, and particle 
velocity $v$ = 0.1). One can observe isolated, but coherently moving flocks with a 
characteristic steady state size coexisting with a background gas of particles.}
\label{snapshot_rec_v0.1}
\end{figure}

As shown in Fig. 3, the PDF was one humped, signaling a second order phase
transition in accord with the earlier results of \cite{vicsek}.
Furthermore, we also determined the corresponding Binder-cumulant $G$, defined above. We found that
$G$ did not exhibit a significant minimum,
corroborating the second order nature of the phase transition. 
On the other hand, the density waves, described by Ref. \cite{gregoire} in the large velocity 
regime, did not occur in the small velocity regime for tractable system sizes. 

\begin{figure}[t!]
\centerline{\resizebox{0.6\textwidth}{!}{\rotatebox{0}{\includegraphics{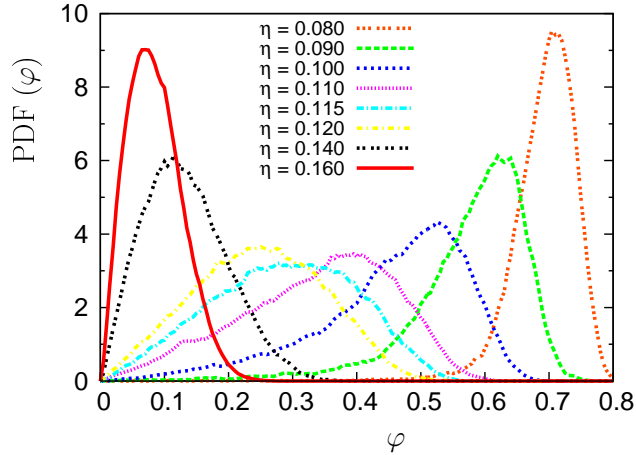}}}}
\caption{ The PDF of the order parameter $\varphi$ in the small velocity regime
($v$ = 0.1) for noise values around the critical point. The one humped character
of the curves demonstrates a {\it second order phase transition}. The curves were
obtained for systems with linear system size $L$ = 512, particle density 
$\rho$ = 1/8 and we used $\simeq100\tau$ MC simulation steps after observing 
the $\tau\simeq 10^5$ relaxation time. }
\label{PDF_v0.1}
\end{figure}

\section{The large velocity regime: boundary condition induced symmetry breaking and density waves} 

Our numerical studies showed that the density waves appear only in
the ordered phase of the large velocity regime ($v\ge$ 0.3). 
In order to elucidate 
the emergence and nature of the density waves, we first 
determined the directional distribution
of the average velocity in the ordered phase when density waves 
are present and found a strong 
anisotropy as shown in Fig. 4. It means that the density waves 
travel mainly parallel to sides of the simulation box or in other cases
they travel in a diagonal direction. This, in fact
implies that {\it the periodic boundary conditions 
have a strong influence on the origin and behavior of the density waves.}

\begin{figure}[t!]
\centerline{\resizebox{0.8\textwidth}{!}{\rotatebox{0}{\includegraphics{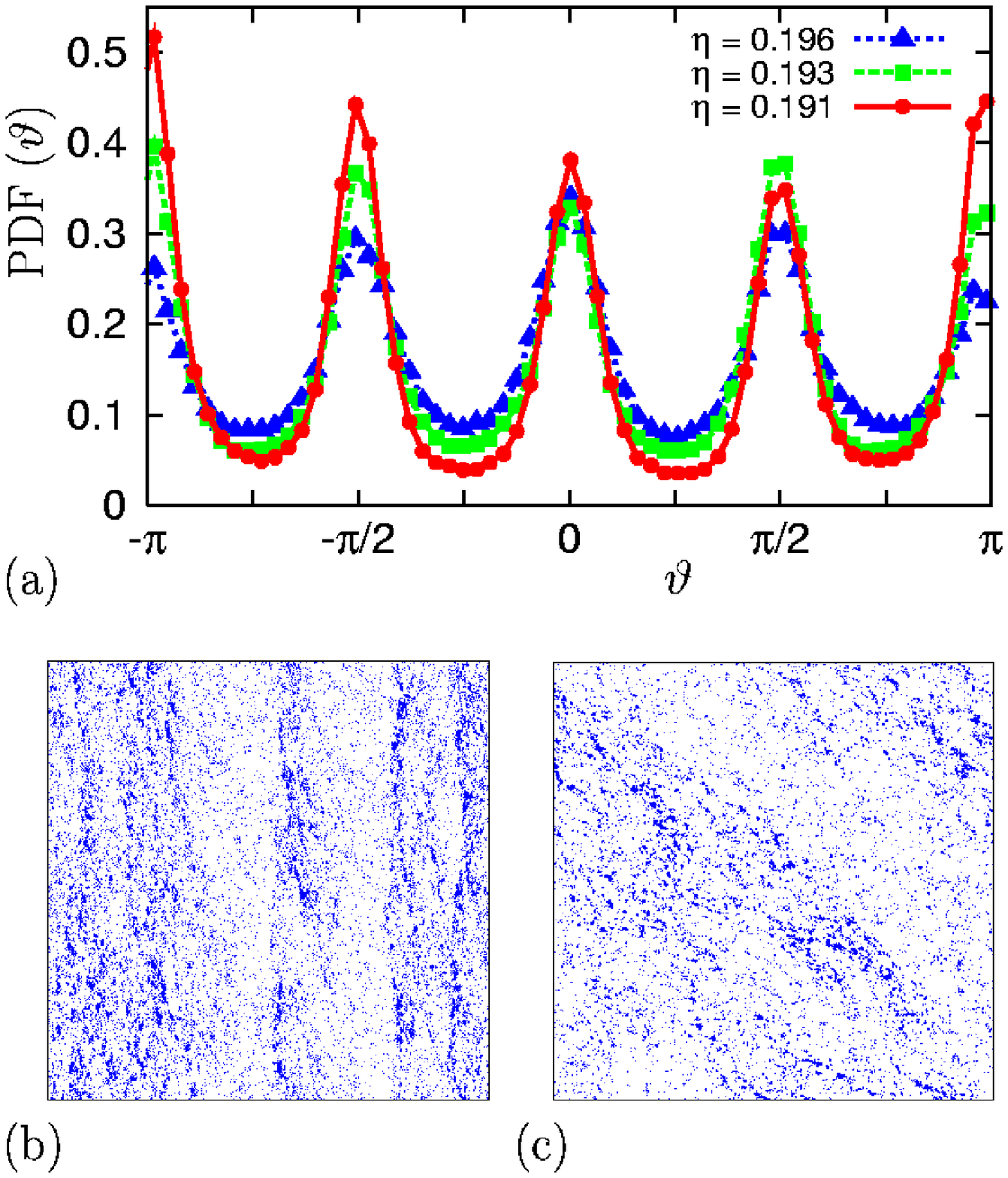}}}}
\caption{ (a) The directional distribution of the average velocity in the large 
velocity regime ($v$ = 0.5) for the 
ordered state at three different values of noise. There is a strong anisotropy, 
i.e., particles preferentially move parallel to the sides of the simulation
box (b) ($v$ = 3), or in other cases they move diagonally (c) ($v$ = 1). 
($L$ = 512, $\rho$ = 1/8.)
}
\label{PhiDir_snapshots}
\end{figure}

To clarify this issue further we implemented a hexagonal simulation cell with 
hexagonal boundary conditions that display a threefold symmetry. 
We found that the directional distribution of the density waves
followed the underlying symmetry of the boundary conditions as demonstrated 
in Fig. 5. Needless to say that such strong influences of the boundary
conditions obscure the physical features and behavior in the large velocity
regime of the SNM. Thus, it is impossible to gain a physical insight or
to draw any conclusions in this regime.  
  
\begin{figure}[t!]
\centerline{\resizebox{0.9\textwidth}{!}{\rotatebox{0}{\includegraphics{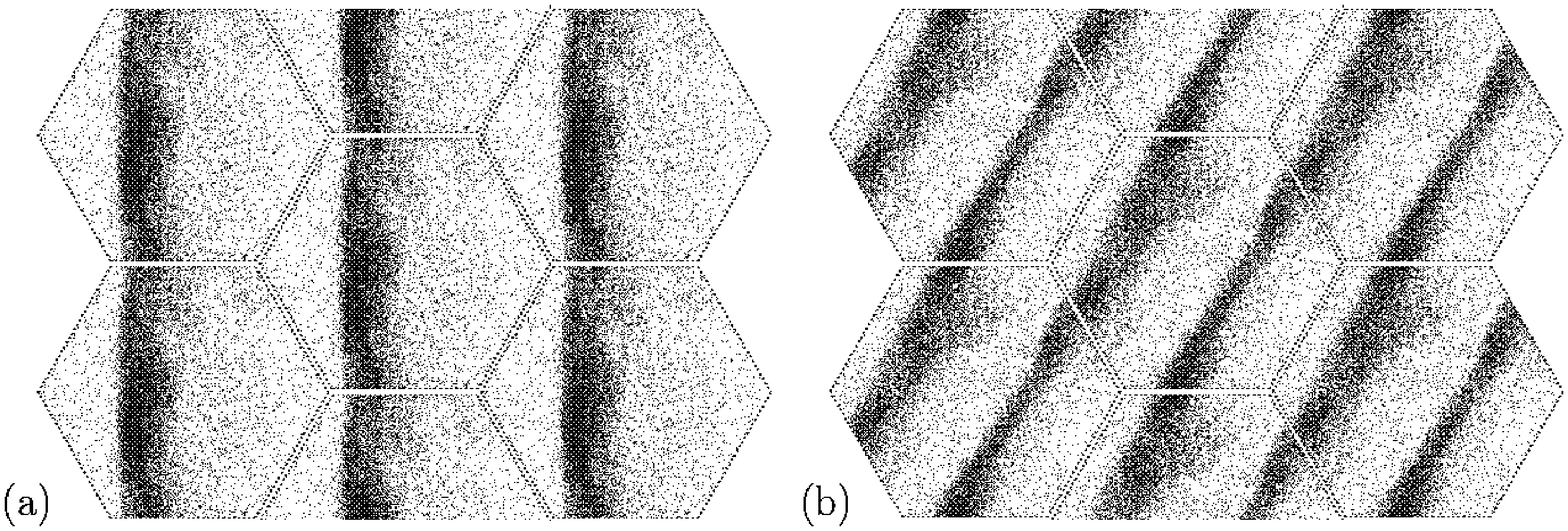}}}}
\caption{Snapshots of the spatial distribution of particles in the 
SNM for a hexagonal simulation cell. The
central cell is repeated to demonstrate the implemented periodic boundary 
conditions (PBC). (a) A density wave traveling in one of the principal directions
of the threefold hexagonal symmetry. (b) A density wave traveling
in another principal direction of the hexagonal system. We found that the 
density waves move in directions determined by the underlying symmetry of
the simulation cell in the large velocity regime ($v\ge$ 0.3). This directional
quantization is a numerical artifact introduced by the presence of the 
PBC. ($L$ = 128/$\sqrt{3}$, $\rho$ = 2/$\sqrt{3}$, $v$ = 10.) }
\label{snapshots_hex}
\end{figure}

In spite of the above discussed principal limits of the physical 
interpretation 
in the large velocity regime, we also investigated this regime numerically.   
Our simulations indicated that at large particle 
velocities (e.g., $v$ = 10) the disorder-order transition exhibits
a discontinuous order parameter and also a negative minimum 
in the Binder cumulant appears that are characteristic features of a first 
order phase transition (Fig. 6). Even though, in the light of the above 
it is not possible to justify this result on a physical basis, we 
note that a similar, lattice version of self-propelled particle models
(Ref. \cite{csahok}) also exhibits a first order phase transition. 
We suspect that the discontinuous phase transition behavior in both 
cases can be attributed to the broken and lowered symmetry in the
system. 

Furthermore, our numerical simulations showed that the phase transition
became of second order again at extreme particle velocities ($v$ = 1000)
in accord with a continuous mean-field like behavior. 
Thus, we witness that the nature of the disorder-order phase transition
changes twice as a function of the particle velocity. The second
order phase transition in the small velocity regime ($v\le$ 0.1)
is replaced by a first order transition like behavior (due to the 
periodic boundary condition induced unphysical symmetry breaking) for large
particle velocities ($v\ge$ 0.3) and the phase transition is
again of second order at extreme particle velocities. We emphasize though, 
that the apparent behavior can physically be justified only in
the small velocity regime ($v\le$ 0.1). 

\begin{figure}[t!]
\centerline{\resizebox{0.9\textwidth}{!}{\rotatebox{0}{\includegraphics{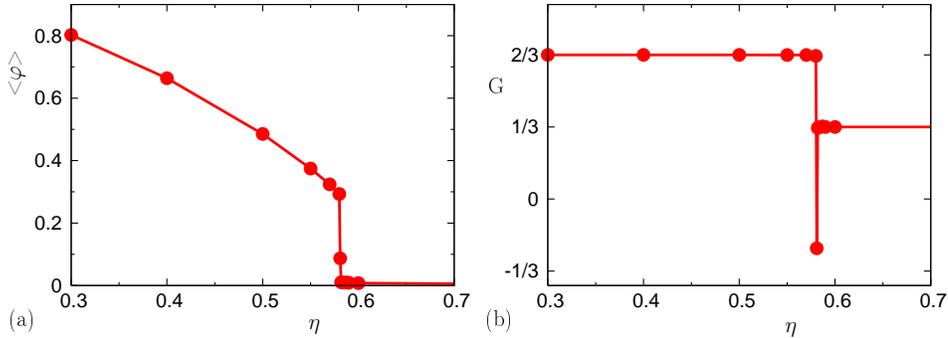}}}}
\caption{ (a) The order parameter $\varphi$ as a function of the noise 
for a very large particle velocity ($v$ = 10). The discontinuous behavior of 
the order parameter indicates a {\it first order phase transition}. 
(b) The corresponding Binder cumulant $G$ displays a sharp and negative minimum
at the phase transition point, characteristic to a first order transition. 
($L$ = 256, $\rho$ = 1, and simulation time $10^7$ MC steps.)
}
\label{phi_binder_v10}
\end{figure}

\section{Particle diffusion} 

In order to further investigate the SNM 
we studied diffusivity of the particles. To that purpose
we took initially neighboring particles and determined their relative 
displacement parallel and perpendicular to the average velocity. 
To avoid the influence of the periodic boundary conditions, 
we considered only relative distances smaller than 1/10 of the linear
system size $L$. Typical
averaged square displacement curves as a function of the time can be seen in
Fig. 7. We find a superdiffusive behavior at intermediate diffusion times in 
the ordered state: $<r^2>(t) \sim t^\alpha$ with $\alpha>$ 1. Furthermore, 
the diffusion is isotropic for small velocities and it becomes anisotropic at
large particle velocities. 

\begin{figure}[t!]
\centerline{\resizebox{0.9\textwidth}{!}{\rotatebox{0}{\includegraphics{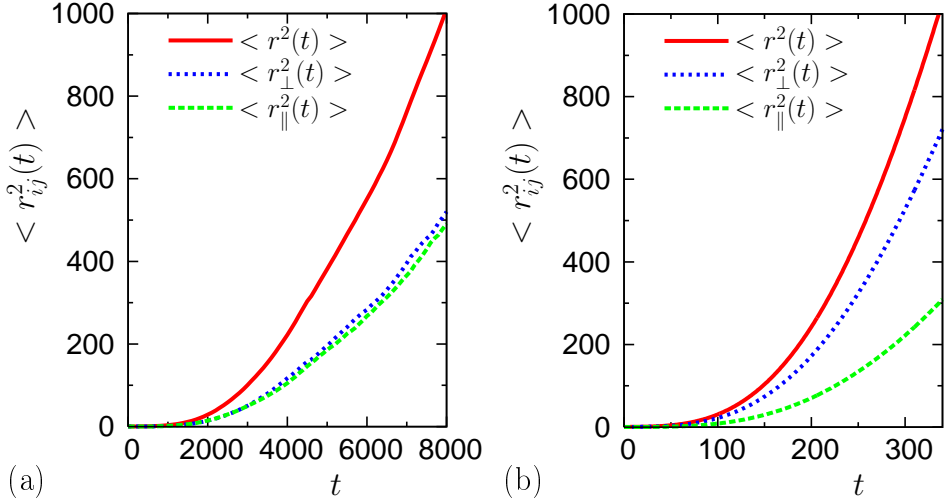}}}}
\caption{ Averaged relative square displacements of initially neighboring
particles as a function of time in the ordered phase ($\eta$ = 0.1). 
(a) The diffusion is isotropic in the small velocity regime ($v$ = 0.05), 
i.e., the relative square displacements are equal in the perpendicular
and parallel directions relative to the average velocity. 
(b) At large particle velocities ($v$ = 0.5) the diffusion is anisotropic, 
i.e., the relative square displacement in the perpendicular direction
($<r^2_\bot>$) is significantly larger than that of in the parallel
direction ($<r^2_\Vert>$). The curves also demonstrate the superdiffusive 
behavior at intermediate diffusion times: $<r^2>(t) \sim t^\alpha$ with
$\alpha>$ 1. ($L$ = 256, $\rho$ = 1/8 and simulation time $10^6$ MC steps.)}
\label{pairDist2}
\end{figure}

The ratio of the square displacement components $A=r^2_\bot/r^2_\Vert$
measures the anisotropy of diffusion. Fig. 8 shows the anisotropy $A$
as a function of the velocity magnitude. At the velocity magnitude
$v$ = 0.3 one can see a remarkable
crossover from the isotropic diffusion to a strongly anisotropic one. 
Thus, while the diffusion is isotropic in the small velocity regime, 
it becomes highly anisotropic for larger velocities. 
We believe that the interplay between the anisotropic diffusion 
for larger velocities and the presence of the periodic boundary 
condition might be responsible for the emergence of the density waves
in the large velocity regime. 

\begin{figure}[t!]
\centerline{\resizebox{0.5\textwidth}{!}{\rotatebox{0}{\includegraphics{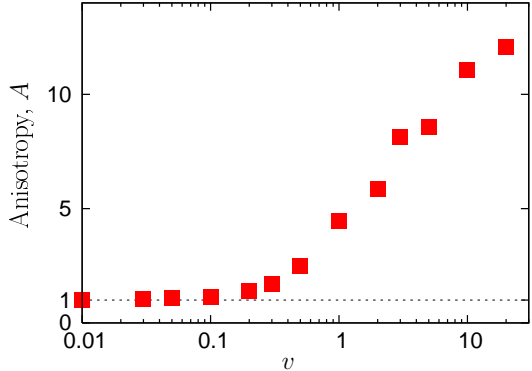}}}}
\caption{ Anisotropy of the particle pair diffusion as a function of the
particle velocity $v$: $A=r^2_\bot/r^2_\Vert$. One can clearly distinguish
the small velocity regime ($v\le$ 0.1) in which the diffusion is isotropic 
from the large velocity regime ($v\ge$ 0.3) with a strongly anisotropic 
diffusion. ($L$ = 256, $\rho$ = 1/8 and simulation time $10^6$ MC steps.)}
\label{anisotropy}
\end{figure}


%

We also measured the 
diffusion of initially neighboring particles as a function of noise
strength. 
The relative mean square displacements of particles are presented in Fig. 9. 
At low noise levels one can distinguish three diffusion regimes. 
At small diffusion times, the 
diffusion process is almost frozen, particles keep their position relative to
each other ($\alpha$ = 0). 
At intermediate times we witness superdiffusion ($\alpha >$ 1). 
Finally, at larger diffusion times, normal diffusion is recovered
($\alpha$ = 1). In order to characterize the extent of superdiffusion
we measured the maximal diffusion exponent $\alpha_{max}$
by determining the maximal slope of the averaged
logarithmic square displacement curves (Fig. 9).  
The noise dependence of the maximal diffusion exponent is plotted in Fig. 10. One can see that 
$\alpha_{max}$ decreases with increasing noise strength and reaches unity at
about the noise strength $\eta$ = 0.5. We believe that this behavior is the 
consequence of the decay of flocks with increasing noise. 
Superdiffusion as well as crossover to normal diffusion were also observed 
and described in Refs. \cite{gregoire,gregoiretu1} for similar models 
of collective motion.  

\begin{figure}[t!]
\centerline{\resizebox{0.6\textwidth}{!}{\rotatebox{0}{\includegraphics{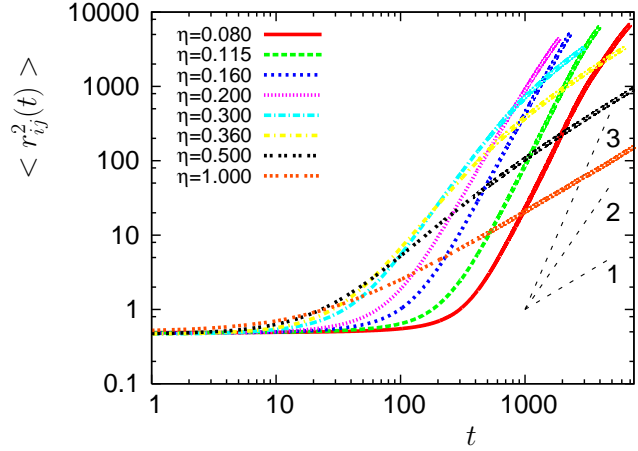}}}}
\caption{ The averaged relative square displacements of initially neighboring
particles $<r^2_{ij}>$ as a function of time for different noise levels. 
Three types of behavior can be observed for low noise levels ($\eta <$ 0.5): 
At small diffusion times no relative
diffusion occurs, at intermediate times superdiffusion dominates and at larger
times normal diffusion is recovered. For high noise levels there is no superdiffusion
due to the lack of coherently moving particle flocks. Three reference
lines with slopes 1, 2, and 3 are also plotted in the figure for visual guidance. 
($L$ = 512, $\rho$ = 1/8, $v=$ 0.1.)}
\label{pairDist_vs_noise}
\end{figure}

\begin{figure}[t!]
\centerline{\resizebox{0.6\textwidth}{!}{\rotatebox{0}{\includegraphics{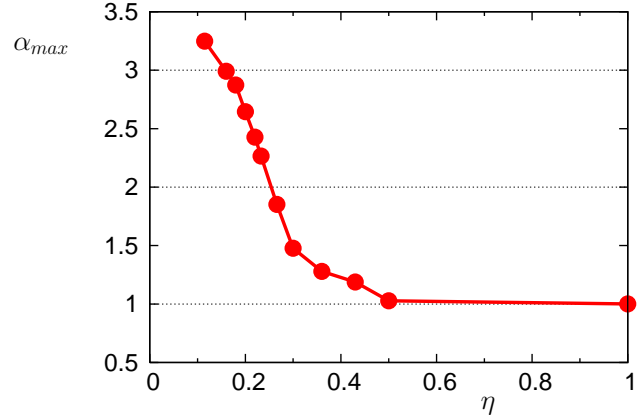}}}}
\caption{
 The noise dependence of the maximal diffusion exponent $\alpha_{max}$. 
This curve was obtained by determining the maximal slope of the averaged
logarithmic square displacement curves (Fig. 9). One can see in accord
with Fig. 9 that there is no superdiffusion above a critical noise value 
($\eta_c\simeq$ 0.5). 
}
\label{pairDist_vs_noise_alpha}
\end{figure}

We interpret the above behavior for small noise strengths as follows. 
At short diffusion times, the majority of initially neighboring 
particles stay together in coherently moving and locally ordered flocks 
(see e.g., Fig. 2), thus, neighboring particles keep their
positions relative to each other ($\alpha$ = 0). At intermediate times, 
i.e., times long enough for particles to change flocks for a few times, 
initially neighboring particles are carried away from 
each other in an ordered manner by the different flocks they belong to. 
It means that the mean spatial separation of the particle pairs 
increase roughly linearly
with time and leads to a superdiffusion exponent ($\alpha \ge$ 2).
At large diffusion times, flocks themselves perform random walks resulting
in a normal diffusion of particle pairs ($\alpha$ = 1). 

Finally, we studied the velocity dependence
of the maximal diffusion exponent $\alpha_{max}$. 
Our results show that $\alpha_{max}$ varied 
between 1.5 and 3 in the superdiffusion time regime (Fig. 11). 

\begin{figure}[t!]
\centerline{\resizebox{0.6\textwidth}{!}{\rotatebox{0}{\includegraphics{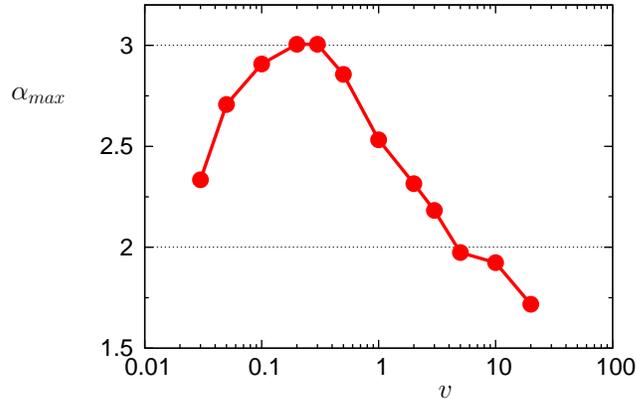}}}}
\caption{ The velocity dependence of the maximal diffusion exponent
$\alpha_{max}$ at a low noise level ($\eta$ = 0.1). This curve shows that 
superdiffusion
is always present below a critical noise value and crossover time. 
($L$ = 256, $\rho$ = 1/8)}
\label{pairDist_vs_vel_alpha}
\end{figure}

\section*{Conclusions} 

In this paper we presented our detailed investigations on the
scalar noise model of collective motion. We justified the
physical relevance of the small velocity regime ($v\le$ 0.1)
and performed extensive numerical simulations 
to re-investigate the order of the order-disorder phase transition
in that regime. 
Our results corroborated the findings of Refs. \cite{vicsek,czirok}, 
i.e., {\it the second order nature of the phase transition}. 

Furthermore, our numerical study of the large velocity regime 
($v\ge$ 0.3) demonstrated the strong effects of the 
boundary conditions on the solutions. We found that the interplay between 
the anisotropic diffusion and the periodic boundary conditions 
introduced a numerical artifact,
the directional quantization of the density waves. 
Thus, the presence of the boundary conditions makes it impossible 
to draw any conclusion about the physical behavior in the large velocity 
regime of the scalar noise model.

Finally, our detailed investigations indicate that the diffusion behavior of 
initially neighboring particles
separates the small and the large velocity regimes. We find that
the diffusion is isotropic for small particle velocities ($v\leq 0.1$), 
it becomes strongly anisotropic at large particle velocities ($v\geq 0.3$). 
Further investigations on the particle pair diffusion revealed 
particle superdiffusion with a crossover to the normal diffusion regime. 
The observed diverse behaviors demonstrate the richness of physics 
exhibited by this simple model.  

This work has been supported by the Hungarian Science Foundation
(OTKA), grant No. T049674 and EU FP6 Grant "Starflag".

\end{document}